\documentclass[aps,pra,reprint, amsmath, amssymb,superscriptaddress,longbibliography]{revtex4-1}
 
\usepackage{bm}
\usepackage[retainorgcmds]{IEEEtrantools}
\usepackage{graphicx}
\usepackage{mathrsfs}
\usepackage{color}
\usepackage{times,txfonts}
\usepackage{nicefrac}
\usepackage{ragged2e}
\usepackage{tikz}
\usepackage{physics}

\usepackage[colorlinks=true,linkcolor=blue,urlcolor=blue,citecolor=blue,pdfusetitle]{hyperref}

\usepackage{blindtext}

\newcommand{\e}{\mathrm{e}}
\newcommand{\diff}{\mathrm{d}}
\newcommand{\I}{\mathrm{i}}
\newcommand{\diag}{\mathrm{diag}}
\newcommand\mydots{\hbox to 1em{.\hss.\hss.}}

\newcommand{\appsection}[2]{\setcounter{equation}{0}\setcounter{subsection}{0}
\section*{Appendix #1. #2}
\renewcommand{\theequation}{#1.\arabic{equation}}
              \renewcommand{\thesection}{#1} }
\catcode`\@=11
\def\numberbysection{\@addtoreset{equation}{section}
        \def\theequation{\thesection.\arabic{equation}}}
\numberbysection

\begin{document}

\title{Exact solution to a Bhatnagar-Gross-Krook-type equation for quantum lattice \\gases with dephasing noise}
%\date{\today}
\author{Michele Coppola}
\email{michele.coppola@ijs.si}
\affiliation{Jožef Stefan Institute, SI-1000 Ljubljana, Slovenia}

\begin{abstract}
The mean-state dynamics of free fermions subject to random projective measurements of local occupation number operators is governed by a Lindblad equation with dephasing noise. In the continuum limit, the equation of motion for the correlation matrix is mapped to a kinetic equation for the Wigner function, which corresponds to a special case of the Bhatnagar–Gross–Krook (BGK) equation without energy conservation. Our main result is the solution to the kinetic equation, showing that the Wigner dynamics emerges from stochastic sampling of classical run-and-tumble processes. As an application, we recover the crossover between ballistic and diffusive transport regimes.
\end{abstract}

\maketitle{}

\section{Introduction}
The measurement postulate is fundamental to quantum mechanics and continues to inspire ongoing debates regarding its interpretation~\cite{von2018mathematical, wheeler2014quantum,wigner1963problem,maudlin1995three,bassi2013models}. Beyond foundational considerations, quantum observables are represented as Hermitian operators acting on the system's Hilbert space, and the measurement process leads to a nondeterministic collapse of the wavefunction. This postulate brings out the invasive nature of quantum measurements, which project the state into the eigenvariety of the corresponding measurement outcome. 

Continuous measurement protocols have recently emerged as ideal platforms for studying the interplay between coherent dynamics and quantum measurements in the spreading of correlations.
These protocols are known for measurement-induced entanglement phase transitions, as local measurements influence the scaling-law behavior of the dynamical and asymptotic von Neumann entropy. Such entanglement transitions have been explored in many different physical settings, e.g., quantum circuits~\cite{skinner2019measurement,bao2020theory,chan2019unitary,li2018quantum,choi2020quantum,szyniszewski2019entanglement,jian2020measurement,lavasani2021measurement,zabalo2020critical,sharma2022measurement,block2022measurement,turkeshi2020measurement}, noninteracting  particles~\cite{chen2020emergent,cao2019entanglement,coppola2022growth,poboiko2023theory,yokomizo2024measurement,lumia2024measurement,alberton2021entanglement,carollo2022entangled,szyniszewski2023disordered,buchhold2021effective,kells2023topological,chatterjee2024measurement}, spin chains~\cite{lang2020entanglement,rossini2020measurement,murciano2023measurement,piccitto2022entanglement,sierant2022dissipative,turkeshi2021measurement,yang2023entanglement,weinstein2023nonlocality}, or Bose-Hubbard type models~\cite{fuji2020measurement,goto2020measurement,doggen2022generalized}. 
For hopping spinless fermions subject to continuous measurements of local occupation number operators, entanglement showcases a single area-law phase in the thermodynamic limit, for both weak and projective measurements~\cite{cao2019entanglement,coppola2022growth,poboiko2023theory}. This area-law behavior is captured by the collapsed quasiparticle pair ansatz~\cite{cao2019entanglement}, where entanglement is carried by pairs of quasiparticles spreading ballistically in opposite directions. Measurements induce random collapse of these pairs, halting their contribution to entanglement growth, and generate new pairs with momentum uniformly distributed in the Brillouin zone. However, this ansatz fails to capture the entanglement dynamics and the entanglement fluctuations~\cite{carollo2022entangled,coppola2022growth}, since the competition between coherent dynamics and random projective measurements also generates multipartite quantum correlations which are inconsistent with an entangled-pair structure, then requiring larger multiplets. Remarkably, the area-law phase also emerges from promising new theoretical approaches based on the Keldysh path integral formalism and the replica trick~\cite{poboiko2023theory}, which could mark a turning point in the study of measurement-induced entanglement phase transitions.

In this work, we also consider chains of hopping spinless fermions, subject to continuous measurements of local occupation number operators. The measurement events follow a Poisson process, giving rise to distinct quantum trajectories, each specified by $N$ measurement outcomes at times $[t_1, \dots, t_N]$. The evolution of the mean state over quantum trajectories is governed by a Lindblad equation with dephasing noise, where the jump operators correspond to local occupation number operators, and the dephasing rate encodes the monitoring frequency. Interestingly, the collapsed quasiparticle pair ansatz was inspired by the generalized hydrodynamics framework~\cite{doyon2020lecture,bastianello2019generalized,capizzi2022domain,scopa2022exact,dubail2017conformal,collura2018analytic,collura2020domain,alba2021generalized,bouchoule2022generalized,bulchandani2017solvable,bulchandani2018bethe,doyon2018soliton,schemmer2019generalized,malvania2021generalized,collura2012entangling,wendenbaum2013hydrodynamic,cao2019entanglement,jin2021interplay,bouchoule2020effect,dast2014quantum,alba2022noninteracting,alba2022hydrodynamics,carollo2022dissipative,bastianello2021hydrodynamics,bastianello2020generalized,fagotti2017higher,fagotti2020locally,dean2019nonequilibrium,moyal1949quantum,scopa2021exact,ruggiero2020quantum,ruggiero2019conformal,coppola2023wigner}, which leads to a kinetic equation for the Wigner function of the mean state. This kinetic equation takes the form of a linear, partial, nonlocal differential equation and corresponds to a special case of the Bhatnagar–Gross–Krook (BGK) equation~\cite{bhatnagar1954model} without energy conservation. The solution to the kinetic equation yields the evolution of the expectation values of quadratic observables, averaged over quantum trajectories. Although many properties of this kinetic equation are well understood, such as the emergent diffusive behavior at long times~\cite{cao2019entanglement,lami2024continuously}, the exact solution is still missing.

In this work, our aim is to derive the solution to the kinetic equation. To achieve this result, we follow the semiclassical approach presented in Ref.~\cite{cao2019entanglement}, commonly referred to as the \emph{quasiparticle picture}. Within this framework, the Wigner function is interpreted as a distribution of classical noninteracting excitations undergoing run-and-tumble processes~\cite{tailleur2008statistical,berg2004coli,patteson2015running,saragosti2012modeling,solon2015active,reichhardt2014active,cates2013active,marchetti2013hydrodynamics,denisov2012levy,zaburdaev2015levy}—a type of stochastic dynamics frequently observed in biological systems and active matter~\cite{tailleur2008statistical,berg2004coli,patteson2015running,saragosti2012modeling,solon2015active,reichhardt2014active,cates2013active,marchetti2013hydrodynamics}. Specifically, run-and-tumble processes model systems alternating between straight-line motion ("run") and random resets in velocity ("tumble"). The nature of these velocity resets is determined by the form of the collision integral in the BGK-type equation, which does not conserve energy. This contrasts with collision integrals for weakly disordered metals, where scattering events with impurities—static objects with spherically symmetric potentials—are typically treated as elastic processes~\cite{mahan2013many,rammer2011quantum,kamenev2023field}.

The motivation behind this work stems from three main considerations. First, the exact solution to the kinetic equation complements the results of Ref.~\cite{cao2019entanglement}, and further validates the quasiparticle description. Second, although our primary focus is on continuously monitored systems, in the continuum limit the same BGK structure characterizes any chain of hopping particles subject to equivalent dephasing noise~\cite{coppola2023wigner}, thereby describing more general dynamics. Third, although a formal solution to this dephasing dynamics—which can be mapped to a Hubbard model with an imaginary interaction strength~\cite{medvedyeva2016exact}—has been obtained via Bethe ansatz techniques, the kinetic equation provides a more intuitive and accessible framework for describing the evolution of quadratic observables. 

The remainder of the paper is organized as follows. In Sec.~\ref{sec:gen}, we define the dynamical protocol and derive the equation of motion for the mean state, which takes the form of a Lindblad equation with homogeneous dephasing noise. In Sec.~\ref{sec:hydro}, we present the kinetic equation and the quasiparticle picture based on run-and-tumble processes. In Sec.~\ref{sec:propagator}, we express the general solution to the kinetic equation in terms of the Wigner propagator. In the quasiparticle picture, the propagator is defined by a scattering probability distribution. This distribution obeys a balance equation that is solved analytically. In Sec.~\ref{sec:transport}, we recover the crossover between ballistic and diffusive transport regimes, previously observed through numerical methods~\cite{cao2019entanglement} and individual random trajectory-based techniques~\cite{lami2024continuously}. Our main findings are summarized in Sec.~\ref{sec:disc}, where we draw some future perspectives. The appendix contains technical details on the solution to the balance equation.

\section{General framework}
\label{sec:gen}
Let us consider a chain of hopping spinless fermions with Hamiltonian
\begin{equation}\label{Hamilt_hopping}
    \hat H = - \frac{1}{2}\sum_{x=-L/2}^{L/2-1} \left(\hat c_x^\dag \hat c_{x+1} + \hat c_{x+1}^\dag \hat c_x \right)\,,
\end{equation}
where the $\hat c$'s are fermionic operators and $L$ (even number) is the total number of sites. Under periodic boundary conditions, the Hamiltonian~\eqref{Hamilt_hopping} can be put in the diagonal form $\hat H = \sum_{p\in\mathcal{B}} \epsilon(p) \hat \eta^\dag_p \hat\eta_p$ by the canonical transformation $\hat\eta_p=\frac{1}{\sqrt{L}}\sum_{x=-L/2}^{L/2-1}\e^{-\I px}\,\hat c_x$, where $p=2\pi \,y/L$ is the momentum belonging to the Brillouin zone $\mathcal{B}=\{2\pi \,y/L\,|\,y\in\mathbb{Z},\,y\in[-L/2,L/2-1]\}$ and $\epsilon(p)=-\cos(p)$ is the single-particle eigenvalue.

We assume that the unitary dynamics is perturbed by random projective measurements of local occupation number operators $\hat n_x=\hat c^\dag_x \hat c_x$, with the monitoring rate $\lambda/2$. Let $\hat\rho(t)$ be the mean state at time $t$. To determine the mean-state dynamics, we first apply the Hamiltonian driving to $\hat\rho(t)$ over a time step $\diff t$, and then we perform local measurements with the probability $\lambda \diff t/2$. At leading order in $\diff t$, the time-evolved mean density matrix reads
\begin{align}\label{preLindblad}
    \hat\rho(t+\diff t) \simeq& \;\frac{\lambda\diff t}{2}\sum_{x=-L/2}^{L/2-1} \bigg(\hat n_x\hat{\rho}(t)\hat n_x + (1-\hat n_x)\hat{\rho}(t)(1-\hat n_x)\bigg)\nonumber\\
    &+\bigg(1-\frac{\lambda\diff t}{2}\bigg)\,\hat\rho(t)\,-\I\comm{\hat H}{\hat\rho(t)}\diff t\;,
\end{align}
where $\hat n_x$ and $1-\hat n_x$ are the projectors onto the eigenvarieties corresponding to outcomes $1$ and $0$, respectively. Therefore, the dynamics of $\hat\rho(t)$ is governed by a Lindblad equation with homogeneous dephasing noise,
\begin{equation}\label{Lindblad}
\frac{\diff \hat\rho}{\diff t} = -\I\comm{\hat H}{\hat\rho} + \lambda\sum_{x=-L/2}^{L/2-1} \left(\hat n_x\hat\rho \hat n_x -\frac{1}{2}\{\hat n_x^2,\hat\rho\}\right)\,,
\end{equation}
where $\hat n_x^2=\hat n_x$ and $\lambda$ is the dephasing rate. Eq.~\eqref{Lindblad} describes the mean-state dynamics of systems under continuous measurements. We also remark that such dephasing noise has long been used to study transport phenomena in boundary-driven quantum systems. In particular, the nonequilibrium steady state exhibits diffusive spin current behavior in various settings, including XX spin chains~\cite{vznidarivc2010exact,vznidarivc2013transport} and Heisenberg models~\cite{vznidarivc2010dephasing}.

The Lindblad equation~\eqref{Lindblad} leads to a closed-form differential equation for the correlation matrix $C_{xy}\equiv\langle \hat c^\dag_{y}\hat c_{x}\rangle(t)=\tr\{\hat c^\dag_{y}\hat c_{x}\,\hat\rho(t)\}$, which reads
\begin{equation}\label{correl}
    \frac{\diff C}{\diff t} = - \I \comm{h}{C} - \lambda \left(C - \diag( C)\right)\,,
\end{equation}
where $\diag(C)$ is the diagonal part of the correlation matrix and $h_{xy}=-(\delta^{(L)}_{x+1,\,y}+\delta^{(L)}_{x-1,\,y})/2$ is the coefficient matrix for the hopping terms in Eq.~\eqref{Hamilt_hopping}. For periodic boundary conditions, $\delta^{(L)}_{x,\,y}=1$ if $x\equiv y\,({\rm mod}\,L)$, and $0$ otherwise.

In the continuum limit, Eq.~\eqref{correl} can be mapped to a kinetic equation for the Wigner function~\cite{cao2019entanglement,coppola2023wigner}, which, as we show in this paper, can be solved by using the quasiparticle picture. 

\section{{Kinetic equation}}
\label{sec:hydro}

The dynamics of two-point correlations is provided by the Wigner function,
\begin{equation}\label{wigner_funct}
    n(x,p,t) = \sum_{y}\,\e^{\I py} \,\langle \hat c^\dag_{x+y/2}\hat c_{x-y/2}\rangle (t)\,.
\end{equation}
The Wigner function is a joint quasiprobability distribution, and its marginals coincide with the single-particle densities in real and momentum space. Strictly speaking, the momentum $p$ in Eq.~\eqref{wigner_funct} only takes discrete values in the Brillouin zone and the sum runs over $y$ such that $x\pm y/2$ are integers. However, we assume that the Wigner function $n(x,p,t)$ varies slowly on the microscopic scales, defined by the lattice spacing and the distance $2\pi/L$ between consecutive Fourier modes. Under this assumption, we can take the continuum limit for both position and momentum variables,  such that the Wigner quasiprobability distribution can be treated as a smooth, analytic function over the phase space $(x,p)$. Starting from Eq.~\eqref{correl} for the two-point functions and neglecting higher-order partial derivatives, the Wigner function satisfies the kinetic equation~\cite{cao2019entanglement,coppola2023wigner}
\begin{equation}\label{Wigner_dyn}
    \partial_t n(x,p,t) + v(p) \partial_x n(x,p,t)= \lambda \left(\rho(x,t) - n(x,p,t)\right)\,,
\end{equation}
where $v(p)=\partial_p \epsilon(p) = \sin(p)$ is the group velocity and
\begin{equation}
    \rho(x,t)=\int_{-\pi}^{\pi}\frac{\diff k}{2\pi} \;n(x,k,t)
\end{equation}
is the local particle density. Although we have framed this work in the context of continuously monitored fermionic systems, the same Lindblad equation~\eqref{Lindblad} applied to bosons ($\hat n_x^2\neq\hat n_x$) leads to an identical equation of motion for the correlation matrix, Eq.~\eqref{correl}. Therefore, the kinetic equation~\eqref{Wigner_dyn} also holds for bosons, as shown in Ref.~\cite{coppola2023wigner}.

Notably, Eq.~\eqref{Lindblad} can be interpreted as a special case of the Bhatnagar–Gross–Krook (BGK) equation~\cite{bhatnagar1954model}—a simplified form of the Boltzmann transport equation in kinetic theory—where the collision integral is
\begin{equation}\label{collision_int}
\partial_t n(x,p,t)|_{\rm coll} = \lambda \left(\rho(x,t) - n(x,p,t)\right)\,.
\end{equation} 
In the absence of dephasing ($\lambda=0$), the collision integral~\eqref{collision_int} vanishes and Eq.~\eqref{Wigner_dyn} reduces to the Moyal equation for closed systems, with formal solution~\cite{fagotti2017higher,fagotti2020locally,dean2019nonequilibrium,moyal1949quantum,scopa2021exact,ruggiero2019conformal,coppola2023wigner},
\begin{equation}\label{moyal}
    n(x,p,t)=n(x-tv(p),p,0)\,.
\end{equation}
Eq.~\eqref{moyal} directly implies ballistic transport. In contrast, for $\lambda > 0$, dephasing leads to more intricate dynamics due to the emergence of a nonlocal term, $\rho(x,t)$.

\begin{figure}
    \centering
    \includegraphics[width=\columnwidth]{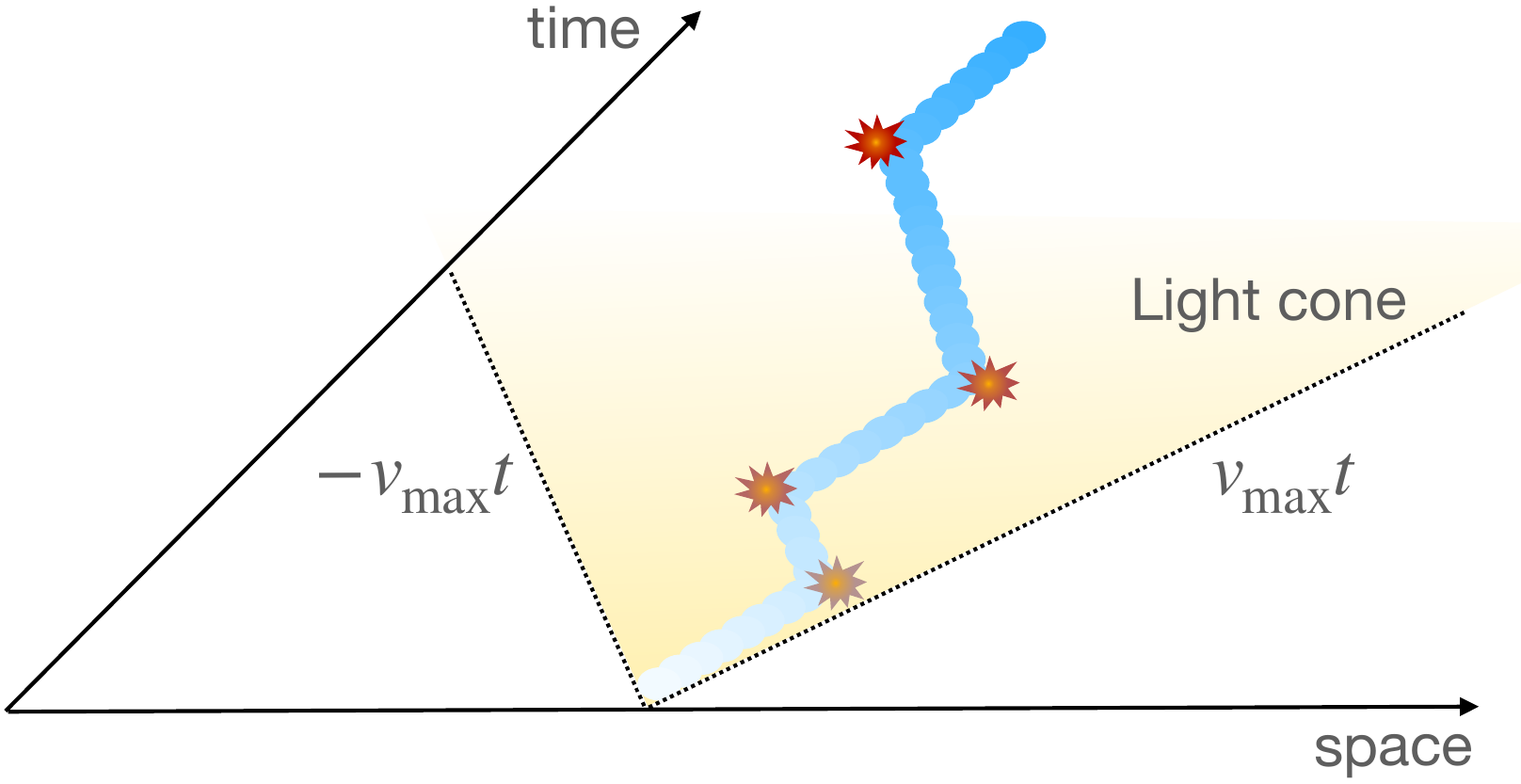}
    \caption{Pictorial representation of a run-and-tumble process: the particle undergoes ballistic flights interrupted by inelastic collisions (represented by red spots) which randomize the particle momentum. The trajectory $x(t)$ is confined within the light cone, i.e. $\abs{x(t)-x(0)}\leq v_{\rm max} t$, where $v_{\rm max} \equiv \max_{p\in\mathcal{B}}\abs{v(p)}=1$ is the maximum group velocity.}
    \label{fig1}
\end{figure}

The main goal of this work is to solve the kinetic equation~\eqref{Wigner_dyn}. To achieve this, we follow the quasiparticle picture in Ref.~\cite{cao2019entanglement}, which has already emerged as a promising ansatz to study transport for continuously monitored systems~\cite{cao2019entanglement,wang2024superdiffusive} or quantum stochastic resistors~\cite{jin2023semiclassical}. The key assumption is that the initial Wigner function $n(x, p, 0)$ can be interpreted as a distribution of classical, noninteracting quasiparticles. After the quench, these excitations propagate at constant velocity $v(p)=\sin(p)$, but for each time step $\diff t$ there is a finite probability $\lambda \diff t$ of picking a new random momentum with uniform distribution in the Brillouin zone $[-\pi,\pi]$. Therefore, the full evolution of the Wigner function $n(x, p, t)$ is obtained as an ensemble average over  classical stochastic trajectories.

As illustrated in Fig.~\ref{fig1}, the quasiparticle motion is an example of a run-and-tumble process~\cite{denisov2012levy,zaburdaev2015levy} where a walker performs a sequence of mutually uncorrelated ballistic flights interrupted by instantaneous collisions. More specifically, the time of flight—the time interval between consecutive scattering events—is a stochastic variable that follows a Poisson distribution, while collisions act as a source of impulsive forces that impart new random momenta to the particles, uniformly distributed within the Brillouin zone. This randomization of momenta has a twofold impact: it makes the collisions inelastic, similar to electron-phonon scattering~\cite{mahan2013many}; it erases any memory of the incoming momenta prior to collisions, leading to uniform transition probabilities between pre- and post-collision momenta. This behavior reflects the effect of dephasing noise, which breaks momentum conservation and delocalize particles in momentum space.

Importantly, as illustrated in Fig.~\ref{fig1}, in a run-and-tumble process, the space-time evolution along any stochastic trajectory is always confined within a light cone, so that at a given time $t$ the walker is always located within the space region $\abs{x(t)-x(0)}\leq v_{\rm max} t$, where $v_{\rm max} \equiv \max_{p\in\mathcal{B}}\abs{v(p)}=1$ is the maximum group velocity.

\section{Wigner dynamics}
\label{sec:propagator}

The kinetic equation~\eqref{Wigner_dyn} is a linear differential equation and the time evolution of the Wigner function over a finite time $t$ can be expressed as an integral kernel, the Wigner propagator $\mathcal{K}(x,p,t|y,k)$, as follows:
\begin{equation}\label{ansartz_propagator}
    n(x,p,t)=\int_{-\infty}^{+\infty}\diff y \int_{-\pi}^{+\pi}\frac{\diff k}{2\pi}\; \mathcal{K}(x,p,t|y,k)\,n(y,k,0)\,,
\end{equation}
where $\mathcal{K}(x,p,t|y,k)$ is the solution of Eq.~\eqref{Wigner_dyn} under the initial condition $\mathcal{K}(x,p,0|y,k)=2\pi\delta(x-y)\delta(p-k)$. In this quasiparticle description, $\mathcal{K}(x,p,0|y,k)=2\pi\delta(x-y)\delta(p-k)$ corresponds to localizing a classical excitation at the phase space point $(y,k)$ at time $t=0$. After the quench, the particle propagates at constant velocity $v(k)=\sin(k)$ until it undergoes inelastic collisions which randomize the momentum, uniformly distributed in $[-\pi,\pi]$. The dephasing noise also makes the time of flight a stochastic variable following the Poisson distribution $\phi(t)=\lambda \e^{-\lambda t}$. The function $\phi(t)$ is a waiting-time distribution, i.e., the probability distribution of detecting the next scattering event after time $t$; $\lambda^{-1}=\int_0^\infty \diff s\,s\,\phi(s)$ is the average waiting time between consecutive collisions. This classical stochastic process represents the starting point for deriving the Wigner propagator. 

Let $\beta(x,t|y,k)$ be the probability distribution of observing a scattering event at position $x$ at time $t$, for a particle initially located at $(y,k)$ at time $t=0$. Following Refs.~\cite{denisov2012levy,zaburdaev2015levy}, $\beta(x,t|y,k)$ satisfies the balance equation
\begin{align}\label{beta}
    \beta(x,t|y,k) = &\int_{-\pi}^{+\pi}\frac{\diff q}{2\pi}\,\int_{0}^{t}\diff s\; \phi(s)\beta(x-s\,v(q),t-s|y,k) \nonumber\\
    &+\phi(t)\,\delta(x-y-t\,v(k))\,,
\end{align}
where $\beta(x-s\,v(q),t-s|y,k)$ is the probability distribution of observing the previous scattering event at position $x-s\,v(q)$ at time $t-s$. The balance equation~\eqref{beta} has a straightforward physical interpretation. The first term accounts for the quasiparticles propagating with velocity $v(q)$ after the last collision at position $x-s\,v(q)$ at time $t-s$. These quasiparticles are the only ones that can scatter again at position $x$ at time $t$, since the motion between consecutive collisions is ballistic. The second term represents the contribution of the quasiparticles undergoing the first scattering event at time $t$.

The solution to the balance equation~\eqref{beta} gives access to the Wigner propagator $\mathcal{K}(x,p,t|y,k)$, which is the probability of finding the excitation at the phase space point $(x,p)$ at time $t$, given that it was initially located at $(y,k)$ at time $t=0$. The evolution of the kernel $\mathcal{K}(x,p,t|y,k)$ is given by 
\begin{align}\label{Propagator}
    \mathcal{K}(x,p,t|y,k) =& \int_{0}^{t}\diff s\; \Phi(s)\beta(x-s\,v(p),t-s|y,k)\nonumber\\
    &+2\pi\,\Phi(t)\,\delta(x-y-t\,v(k))\,\delta(p-k)\,,
\end{align}
where 
\begin{equation}\label{Phi}
    \Phi(s)=1-\int_0^s \diff r\;\phi(r)=\frac{\phi(s)}{\lambda}\,,
\end{equation}
is the no-jump probability, i.e., the probability that no scattering events occur over a time interval of length $s$. The two terms in Eq.~\eqref{Propagator} have distinct origins. The first one accounts for the quasiparticles which undergo a scattering event at position $x-s\,v(p)$ at time $t-s$ and then propagate ballistically with momentum $p$ over the time interval $[t-s,t]$ without further collisions. The second term accounts for the quasiparticles that, after the quench, propagate with momentum $p$ along a deterministic ballistic trajectory, reaching the site $x$ at time $t$ if $y=x-tv(p)$. In fact, in the less probable scenario, exponentially suppressed over time by the Poisson distribution, the particle never collides and propagates with constant velocity $v(p)$. 

The main result of this paper is the analytical solution to the balance equation~\eqref{beta}, which gives access to the Wigner propagator $\mathcal{K}(x,p,t|y,k)$. In the Appendix~A, we prove that the scattering probability distribution is
\begin{align}\label{main_res}
    \beta(x,t|y,k)=&\,\phi(t)\delta(x-y-t\,v(k))\nonumber\\
    &+\int_0^t \diff s\;\phi(s)\,f(x-y-s\,v(k),t-s)\,,
\end{align}
with
\begin{align}
        f(x,t)\equiv&\frac{\lambda e^{-\lambda t}}{\pi\sqrt{t^2-x^2}}\bigg[1+\frac{\pi\lambda}{2}\sqrt{t^2-x^2}\bigg(I_0\left(\lambda\sqrt{t^2-x^2}\right)\nonumber\\
        &+L_0\left(\lambda\sqrt{t^2-x^2}\right)\bigg)\bigg]\Theta(t-\abs{x})\,,
\end{align}
where $I_0$ is the modified Bessel function of the first kind, $L_0$ is the modified Struve function and $\Theta$ is the Heaviside step function.

Although the general solution to the kinetic equation is rather involved, the Wigner dynamics simplifies considerably for initial profiles that are homogeneous in either space or momentum.
In the first scenario where $n(x,p,0)=n(p,0)$, $n(x,p,t)=n(p,t)$ remains space independent during time evolution and
\begin{equation}\label{thermal}
    n(p,t)=(1-\e^{-\lambda t})\int_{-\pi}^{+\pi}\frac{\diff k}{2\pi} n(k,0) \;+\; \e^{-\lambda t} n(p,0)\,.
\end{equation}
For initial thermal states at inverse temperature $\beta$ and zero chemical potential, $n(k,0)=(\e^{\beta\epsilon(k)}+1)^{-1}$, and Eq.~\eqref{thermal} leads to $n(p,t)=(1-\e^{-\lambda t})/2 + \e^{-\lambda t} n(p,0)$, with the asymptotic solution corresponding to an infinite temperature state~\cite{vznidarivc2010exact,cao2019entanglement,coppola2023some}.  Another interesting case is $n(x,p,0)=n(x,0)=\rho(x,0)$, where the local particle density $\rho(x,t)$ can be easily derived. First, Eqs.~(\ref{beta},~\ref{Propagator},~\ref{Phi}) yield
\begin{equation}
    \int_{-\pi}^{+\pi}\frac{\diff p}{2\pi}\; \mathcal{K}(x,p,t|y,k)=\frac{1}{\lambda}\,\beta(x,t|y,k)\,.
\end{equation}
Second, applying the convolution theorem analogously to the Appendix A, we obtain
\begin{equation}
    \int_{-\pi}^{+\pi}\frac{\diff k}{2\pi}\;\beta(x,t|y,k)=f(x-y,t)\,,
\end{equation}
which finally implies
\begin{equation}\label{equazione_rho}
    \rho(x,t)=\frac{1}{\lambda}\int_{-\infty}^{+\infty}\diff y \;f(x-y,t)\,\rho(y,0)\,.
\end{equation}
Therefore, the density profile is simply given by a convolution. For example, for domain wall initial conditions, $n(x,0)=\rho(x,0)=\Theta(x)$, Eq.~\eqref{equazione_rho} reproduces the results in Ref.~\cite{ishiyama2025exact}.
Finally, for homogeneous profiles $n(x,p,0)=n(0)$, we get $n(x,p,t)=n(0)$, which indicates that the state is stationary. 

\section{Nonequilibrium quantum transport}
\label{sec:transport}

As formulated, the collision integral~\eqref{collision_int} conserves the particle number, but not energy or momentum~\cite{bhatnagar1954model}. From Eq.~\eqref{Wigner_dyn}, the particle density $\rho(x,t)$ satisfies the continuity equation 
\begin{equation}\label{continuity_eq}
    \partial_t \rho(x,t) = -\partial_x j(x,t)\,,
\end{equation}
where 
\begin{equation}\label{current}
    j(x,t)=\int_{-\pi}^{\pi}\frac{\diff p}{2\pi} \;v(p)\,n(x,p,t)
\end{equation}
represents the particle current. Moreover, the evolution equation for the current reads
\begin{equation}\label{current_dyn}
    \partial_t j(x,t) + \partial_x \int_{-\pi}^{\pi}\frac{\diff p}{2\pi} \;v(p)^2\,n(x,p,t)=-\lambda j(x,t)\,.
\end{equation}
Building on the quasiparticle picture, collisions randomize particle momenta following a uniform distribution, so that at large times ($t\gg\lambda^{-1}$) these scattering events drive the system toward locally homogeneous Wigner profiles in momentum space, $n(x,p,t)\simeq\rho(x,t)$, and hydrodynamic behavior emerges. After an initial transient ($t\sim\lambda^{-1}$) during which the damping term in Eq.~\eqref{current_dyn} induces fast decay of the current, the particle current settles into
\begin{equation}
    j(x,t)\simeq -\frac{1}{2\lambda}\partial_x\rho(x,t)\,.
\end{equation}
Together with the continuity equation~\eqref{continuity_eq}, this yields Fick’s law,
\begin{equation}\label{heat}
    \partial_t \rho(x,t)\simeq\frac{1}{2\lambda}\partial_{xx}\rho(x,t)\,,
\end{equation}
for diffusive transport. This behavior aligns with earlier works, where a ballistic-to-diffusive transition has been both numerically observed~\cite{cao2019entanglement} and analytically demonstrated~\cite{lami2024continuously}. In this section, we first show that the mean squared displacement aligns with the findings in Refs.~\cite{cao2019entanglement,lami2024continuously}. Then, we extend the results of Refs.~\cite{cao2019entanglement,lami2024continuously} by studying the asymptotic form of the Wigner function.

\begin{figure}
    \centering
    \includegraphics[width=\columnwidth]{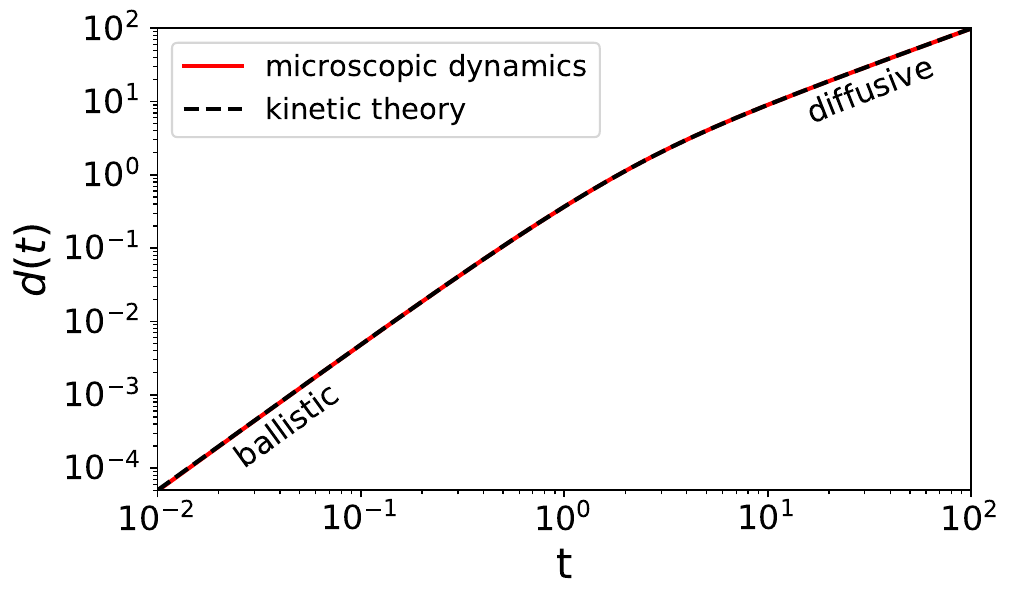}
    \caption{Evolution of the mean squared displacement $d(t)$ for $n(x,0)=\rho(x,0)=\delta(x)$ and $\lambda=1$: kinetic theory~\eqref{displacement} (black dashed line) vs microscopic dynamics~\eqref{correl} (red curve). The crossover between ballistic and diffusive transport regimes occurs at the characteristic time $\lambda^{-1}=1$. As expected, we find strong agreement between numerics and analytical results.}
    \label{fig2}
\end{figure}

Let us prepare the quantum system in the initial state $n(x,0)=\rho(x,0)=\delta(x)$. Thanks to Eq.~\eqref{equazione_rho}, the time-evolved local particle density is $\rho(x,t)=f(x,t)/\lambda$. By definition, the mean squared displacement is $d(t)\equiv\langle x^2\rangle - \langle x\rangle^2$. However, by symmetry $\langle x\rangle(t) = 0$ $\forall t>0$, and thus
\begin{equation}\label{displacement}
    d(t) = \int_{-\infty}^{+\infty}\diff x \;x^2\,\rho(x,t) = \frac{t}{\lambda} + \frac{1}{\lambda^2}\left(\e^{-\lambda t}-1\right)\,.
\end{equation}
Eq.~\eqref{displacement} showcases the expected transition. Indeed, the second moment evolves according to the following scaling behavior,
\begin{equation}\label{trns}
  d(t)\simeq   \begin{cases}
        \;t^2/2\,,\hspace{0.65cm}\lambda t\ll 1\,,\hspace{0.8cm}\text{ballistic regime}\,,\\
        \;t/\lambda\,,\hspace{0.8cm}\lambda t\gg 1\,,\hspace{0.8cm}\text{diffusive regime}\,.
    \end{cases}
\end{equation}
Eq.~\eqref{trns} is in agreement with the numerical findings in Ref.~\cite{cao2019entanglement} and the analytical results in Ref.~\cite{lami2024continuously}. In Fig.~\ref{fig2}, we show the evolution of the mean squared displacement $d(t)$ for $n(x,0)=\rho(x,0)=\delta(x)$ and $\lambda=1$, in a log-log scale. The black dashed line corresponds to the analytical solution~\eqref{displacement}, while the red curve represents the microscopic dynamics, obtained via numerical integration of Eq.~\eqref{correl} using the Euler method, for a system of $L = 100$ sites.

It is also interesting to observe how the ballistic-to-diffusive transition shows up at the level of the Wigner quasiprobability distribution. Under the initial condition $n(x,0)=\rho(x,0)=\delta(x)$, we get
\begin{equation}
      \displaystyle \rho(x,t)\simeq   \begin{cases}
        \;\displaystyle\frac{\Theta(t-\abs{x})}{\pi\sqrt{t^2-x^2}}\,,\hspace{1.1cm}\lambda t\ll 1\,,\\
        \\
        \;\displaystyle\sqrt{\frac{\lambda}{2\pi t}} \e^{-\frac{\lambda x^2}{2t}}  \,,\hspace{1cm}\lambda t\gg 1\,,\;\frac{\abs{x}}{t}\ll 1\,.
    \end{cases}
\end{equation}
As a sanity check, if $\lambda t\ll 1$ we recover the solution for the unitary dynamics. In fact, in the absence of dephasing ($\lambda=0$), Eq.~\eqref{moyal} implies $n(x,p,t)=\delta(x-tv(p))$.
On the other hand, for $\lambda t\gg 1$ and $\abs{x}/t\ll 1$, the local particle density is approximated by a Gaussian distribution, which notoriously obeys Fick's equation~\eqref{heat} for diffusive transport.

\begin{figure}
    \centering
    \includegraphics[width=\columnwidth]{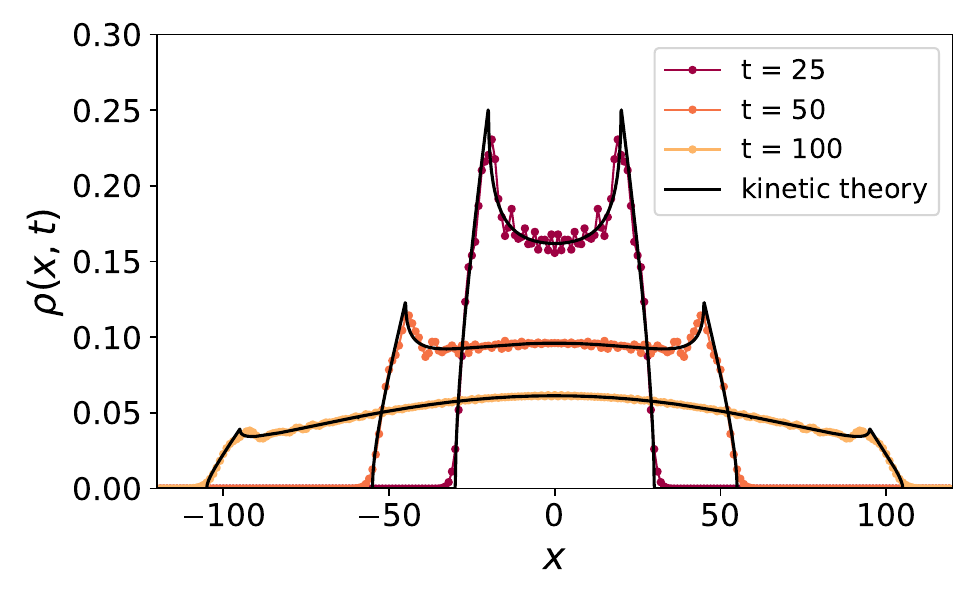}
    \caption{Dynamics of the local particle density: kinetic theory (black lines)~\eqref{equazione_rho} vs microscopic dynamics (colored dots)~\eqref{correl}. We plot the density profiles at times $t\in\{25,50,100\}$ for the initial double domain wall $n(x,0)=\rho(x,0)=\Theta(x+l/2)-\Theta(x-l/2)$, $l=10$, and $\lambda=0.1$.}
    \label{fig3}
\end{figure}

Fig.~\ref{fig3} compares the density profiles obtained from the kinetic theory, Eq.~\eqref{equazione_rho}, and from the microscopic dynamics, Eq.~\eqref{correl}. The latter is computed by numerical integration of Eq.~\eqref{correl} using the Euler method for a system of $L=240$ sites. In Fig.~\ref{fig3}, we prepare the system in a double domain wall, $n(x,0)=\rho(x,0)=\Theta(x+l/2)-\Theta(x-l/2)$, at time $t=0$. However, this initial Wigner profile is not microscopically equivalent to setting $C_{jj}(0)=1$ for $j\in[-l/2,l/2]$ and $0$ otherwise as the initial condition for the correlation matrix. Indeed, due to discretization, the total number of particles differs; specifically, $l=\int\diff x \, \rho(x,0)\neq \sum_{j}C_{jj}(0)=l+1$. To ensure the same total number of particles, here we set $C_{jj}(0)=\frac{l}{l+1}$ for $j\in[-l/2,l/2]$, noting that the offset vanishes in the limit $l\to \infty$, as expected. Fig.~\ref{fig3} shows the evolution of the density profiles at times $t\in\{25,50,100\}$, for $l=10$ and the dephasing rate $\lambda=0.02$. Therefore, time $t=25$ corresponds to the ballistic regime, where we can observe the front waves spreading ballistically. The ballistic-to-diffusive transition occurs at time $t=50$ and at time $t=100$ the system is in the diffusive regime. The agreement between the microscopic dynamics and the kinetic model is good, with small discrepancies arising from the continuum limit and neglected higher-order partial derivatives in the BGK-type equation~\eqref{Wigner_dyn}.

\section{Discussion and conclusion}
\label{sec:disc}
The mean-state dynamics of free fermions subject to random projective measurements of local occupation number operators is governed by a Lindblad equation with dephasing noise. In the continuum limit, the Wigner function satisfies the kinetic equation~\eqref{Wigner_dyn} and is interpreted as a joint probability distribution of classical noninteracting excitations undergoing run-and-tumble processes.
The general solution to the kinetic equation is formulated in terms of the Wigner propagator. In the quasiparticle picture, the propagator is given by the scattering probability distribution, which satisfies the balance equation~\eqref{beta}. As a main result, we provide the analytical solution to the balance equation~\eqref{beta} and, as an application, we recover the ballistic-to-diffusive transition, which occurs at the characteristic time $t=\lambda^{-1}$. Once again, we demonstrate that the diffusive behavior emerges for any finite value of $\lambda$, regardless of how small it is~\cite{ishiyama2025exact}. Specifically, in the regime $\lambda t\gg 1$ and $\abs{x}/t\ll 1$, the local particle density approaches a Gaussian distribution, which obeys Fick's equation~\eqref{heat}. Finally, we compare the density profiles predicted by the kinetic theory with those obtained from the microscopic dynamics, showing good agreement.

This work adds another piece to our understanding of dephasing dynamics and further validates the quasiparticle picture. Moreover, the approach developed in this work can be easily generalized to incorporate momentum-dependent scattering rates, as studied in Ref.~\cite{wang2024superdiffusive}, where similar dephasing dynamics induce superdiffusive transport at long times. This represents an interesting direction for future research.

\section*{Acknowledgments}
I acknowledge the support by the QuantERA II JTC 2021 Grants QuSiED and T-NiSQ by MVZI, the P1-0044 program of the Slovenian Research Agency and ERC StG 2022 Project DrumS, under Grant Agreement No.~101077265. 

I am thankful to Jerôme Dubail, Zala Lenarčič, Mario Collura, and Gianluca Lagnese for fruitful discussions.

\section*{Data availability}
The data that support the findings of this article are openly available~\cite{datamanuscript}.

\appsection{A}{Exact solution to the balance equation}
\label{sec:balance}

In this appendix, we explicitly solve the balance equation~\eqref{beta} which leads to the Wigner propagator $\mathcal{K}(x,p,t|y,k)$ and the full dynamics of the Wigner function $n(x,p,t)$.

First, let us define the Fourier ($\mathcal{F}$) and Laplace ($\mathcal{L}$) transforms in space $(x)$ and time $(t)$ domains, respectively. Let $g_1(x)$ and $g_2(t)$ be two arbitrary functions of space and time. Their transforms are defined as
\begin{equation}
\mathcal{F}[g_1(x)](z)\equiv\int_{-\infty}^{+\infty}\diff x\; \e^{-\I xz}g_1(x)\,,
\end{equation}
\begin{equation}
    \mathcal{L}[g_2(t)](\gamma)\equiv\int_0^\infty \diff t\; \e^{-\gamma t}g_2(t)\,,
\end{equation}
where $z$ and $\gamma$ are the reciprocal variables of space and time, respectively.

We notice that, after each collision, a random quasiparticle momentum is assigned, uniformly distributed within the Brillouin zone $[-\pi,\pi]$. This is analogous to assigning a random quasiparticle velocity $v$ with probability distribution 
\begin{equation}
    \mu(v)=\frac{1}{\pi}\frac{1}{\sqrt{1-v^2}}\,\Theta(1-\abs{v})\,,
\end{equation}
where $\Theta(1-\abs{v})$ is the Heaviside step function.
Therefore, with a change of variables, the balance equation~\eqref{beta} becomes
\begin{align}\label{beta2}
    \beta(x,t|y,k) =& \int_{-\infty}^{+\infty}\diff v\int_{0}^{t}\diff s\; \phi(s)\mu(v)\beta(x-sv,t-s|y,k)\nonumber\\
    &+ \phi(t)\delta(x-y-t\,v(k))\,.
\end{align}
Eq.~\eqref{beta2} can be solved by using the convolution theorem. Let us define $\Tilde\beta(z,\gamma|y,k)\equiv\mathcal{L}\mathcal{F}[\beta(x,t|y,k)]$ as the Laplace-Fourier transform of the scattering probability density $\beta(x,t|y,k)$. One can show that
\begin{equation}
    \Tilde\beta(z,\gamma|y,k) = \frac{\mathcal{L}\mathcal{F}\bigg[\phi(t)\delta(x-y-t\,v(k))\bigg]}{1-\mathcal{L}\bigg[\phi(t)\mathcal{F}[\mu(x)](tz)\bigg]}\,,
\end{equation}
where $\mathcal{F}[\mu(x)](tz)$ is the Fourier transform of the distribution $\mu$ evaluated at $tz$. Since  
\begin{equation}
    \mathcal{L}\bigg[\phi(t)\mathcal{F}[\mu(x)](tz)\bigg] = \frac{\lambda}{\sqrt{z^2+(\lambda+\gamma)^2}}\,,
\end{equation}
we get 
\begin{align}\label{transformed_beta}
    \Tilde\beta(z,\gamma|y,k) =& \mathcal{L}\mathcal{F}\bigg[\phi(t)\delta(x-y-t\,v(k))\bigg]\nonumber\\
    &\times\left[1+\sum_{n=0}^\infty\left[\left(\frac{z}{\lambda}\right)^2+\left(\frac{\lambda+\gamma}{\lambda}\right)^2\right]^{-\frac{n+1}{2}}\right]\,.
\end{align}
At this point, we only need to back transform to $\beta(x,t)$. One can show that
\begin{equation}
    \mathcal{L}^{-1}\mathcal{F}^{-1}\left[\sum_{n=0}^\infty\left[\left(\frac{z}{\lambda}\right)^2+\left(\frac{\lambda+\gamma}{\lambda}\right)^2\right]^{-\frac{n+1}{2}}\right] = f(x,t)\,,
\end{equation}
and applying the convolution theorem to Eq.~\eqref{transformed_beta}, we finally get the scattering probability distribution~\eqref{main_res}.

\bibliography{biblio}

\end{document}